\documentclass[12pt,a4paper,english,preprint,superscriptaddress,aps]{revtex4-1}
\usepackage[utf8]{inputenc}
\usepackage{babel}
\usepackage{amsmath}
\usepackage{amssymb}
\usepackage[unicode=true,
 bookmarks=false,
 breaklinks=false,pdfborder={0 0 1},backref=section,colorlinks=false]
 {hyperref}

\makeatletter

\pdfpageheight\paperheight
\pdfpagewidth\paperwidth

\usepackage{babel}

\makeatother

\begin{document}

\title{Reduction of order and Fadeev-Jackiw formalism in generalized electrodynamics}

\author{A. A. Nogueira}
\email{andsogueira@hotmail.com}

\affiliation{Universidade Federal do ABC, Centro de Ciências Naturais e Humanas,
Rua Santa Adélia, 166, 09210-170, Santo André, SP, Brasil}

\author{C. Palechor}
\email{caanpaip@gmail.com}

\affiliation{Universidade Federal do ABC, Centro de Ciências Naturais e Humanas,
Rua Santa Adélia, 166, 09210-170, Santo André, SP, Brasil}

\author{A. F. Ferrari}
\email{alysson.ferrari@ufabc.edu.br}

\affiliation{Universidade Federal do ABC, Centro de Ciências Naturais e Humanas,
Rua Santa Adélia, 166, 09210-170, Santo André, SP, Brasil}

\affiliation{Indiana University Center for Spacetime Symmetries, Indiana University,
Bloomington, Indiana 47405-7105}
\begin{abstract}
The aim of this work is to discuss some aspects of the reduction of
order formalism in the context of the Fadeev-Jackiw symplectic formalism,
both at the classical and the quantum level. We start by reviewing
the symplectic analysis in a regular theory (a higher derivative massless
scalar theory), both using the Ostrogradsky prescription and also
by reducing the order of the Lagrangian with an auxiliary field, showing
the equivalence of these two approaches. The interpretation of the
degrees of freedom is discussed in some detail. Finally, we perform
the similar analysis in a singular higher derivative gauge theory
(the Podolsky electrodynamics), in the reduced order formalism: we
claim that this approach have the advantage of clearly separating
the symplectic structure of the model into a Maxwell and a Proca (ghost)
sector, thus complementing the understanding of the degrees of freedom
of the theory and simplifying calculations involving matrices. 
\end{abstract}
\maketitle

\section{Introduction\label{sec1}}

Constrained systems are a basic tool for theoretical research in different
contexts such as gauge theories and the field theory approach for
gravity, for example. The pioneers in this treatment were Dirac and
Bergmann (DB) \citep{Dirac,Bergmann,Diracbook} whose works established
the standard method to study theories with constraints, providing
generalized brackets appropriate to quantize these systems. When the
dynamics of a singular Lagrangian formulated in configuration space
is translated to a Hamiltonian formulation in phase space, the first
constraints that appear, from the definition of the canonical momenta,
are called the Dirac primary constraints. The condition that these
should not change over time (consistency condition) may generate additional
constraints, called secondary constraints, for which consistency conditions
are again applied, and so forth. From this iterated process we obtain
a complete set of constraints, which we may classify as being of first
or second class, according to the vanishing or not of their canonical
Poisson brackets. This Dirac-Bergmann algorithm, including its classification
of constraints, has a meaning associated with the physical degrees
of freedom \citep{SundRothe}. This provides a first approach to the
connection between classical and quantum dynamics, the classical dynamics
described in the phase space by the observables and (Poisson$\backslash$Dirac)
brackets, and the quantum dynamics described in Hilbert space by the
operators and commutator\textbackslash anti-commutators. A second
approach begins in a study by Dirac about the connection between a
classical dynamics described in configuration space and its resulting
quantum description, where we see the emergence of a very important
object called the transition amplitude \citep{Dirac2}. Feynman later
used Dirac's idea to describe the quantum Lagrangian mechanics with
the path integral formalism \citep{Feyn}; afterwards, an elegant
variational principle \citep{Noether} of the quantum action was developed
by Schwinger, utilizing as a guide the Heisenberg description \citep{Schwinger}.

The need to describe the interactions of nature along the lines of
a relativistic dynamics leads us to build a covariant language with
gauge symmetry \citep{Ut}, which has more degrees of freedom then
the physical ones, hence the necessity of introducing constrains.
The connections between classical and quantum physical systems with
constraints, in a functional formalism, was first formulated by Faddeev
(for first class constraints), and later extended by Senjanovic (including
second class constraints) \citep{Fadd}. The quantization procedure
of a gauge theory is in principle possible for the physical degrees
of freedom only, and thus we loose the explicit covariance of the
equations: in order to maintain it at the quantum level, Faddeev,
Popov and DeWitt built a method in which additional, non-physical
ghost fields, are introduced \citep{faddeev}.

The canonical quantization gained new life with the Fadeev-Jackiw
(FJ) method, developed in the 1980's \citep{FJ}. The (FJ) formalism
pursues a classical geometric treatment based on the symplectic structure
of the phase space and it is only applied to first order Lagrangians.
The 2-form symplectic matrix associated with the reduced Lagrangian
allows us to obtain the generalized brackets in the reduced phase
space without the need to follow Dirac's method step by step \citep{Jackiwtears}.
The (FJ) method has some very useful properties, such as not needing
to distinguish the types of constraints and the Dirac's conjecture,
and therefore evoked much attention. Barcelos and Wotzasek introduced
one procedure of dealing with constraints in the (FJ) method \citep{BW,MW};
on the other hand, despite the quantization being essentially canonical,
the path integral quantization was also constructed in \citep{Li1,Toms}.
We can find in the literature many studies of the equivalence between
the (DB) and (FJ) formalisms \citep{Li2,Li3,Jairzinho,Zambrano},
which can be proved in many (but not all) cases.

When Ostrogradski constructed Lagrangian theories with higher order
derivatives in classical mechanics, a new field of research was opened
\citep{Ostro,Pomp}. Bopp, Podolsky and Schwed \citep{Bop} proposed
a generalized electrodynamics in an endeavor to get rid of the infinites
in quantum electrodynamics (QED), starting from a higher order Lagrangian,
corresponding to the usual QED Lagrangian augmented by a term quadratic
in the divergence of the field-strength tensor, which by dimensional
reasons introduces a free parameter that can be identified as the
Podolsky's mass $m$. This modification gives the correct (finite)
expression for the self-force of charged particles, as shown by Frenkel,
and interesting effects produced by the presence of external sources
\citep{Frank,FGA2}. At the quantum level, higher derivative theories
have in general the property of better behaved (or even absent) ultraviolet
divergences in a sense closely related to the Pauli-Villars-Rayski
regularization scheme \citep{PV,Rays}, but also sometimes exhibit
Hamiltonians without a lower limit \citep{PU} due to the presence
of states with negative norm (ghosts), leading to the breakdown of
unitarity \citep{WH}. Several procedures to avoid this problem have
been already been studied \citep{Hawk}, one approach being a careful
investigation of the analytic structure of the Green functions as
discussed in \citep{Stelle,Shapiro,Anselmi}. Another way to implement
terms with higher order derivatives without breaking the stability
of the theory has recently been proposed using the concept of Lagrangian
anchors \citep{Russos}, an extension of the Noether theorems in the
sense that one defines a class of conserved quantities associated
with a given symmetry. For instance, the symmetry due to time translations
will lead to two conserved Hamiltonians, one of them will have regularizing
properties but will break the stability because the energy is not
bounded from below, whereas the other recovers the stability but loses
the regularizing properties. This leads to a new perspective on the
unitarity problem of higher derivative theories, that makes use of
the formalisms of reduction of order \citep{Reduced,Red} and the
concepts of complexation of the Lagrangian \citep{Complex}.

Given the advantages in using the (FJ) formalism to deal with the
constraint structure of gauge theories, it is natural to apply this
formalism to the quantization of gauge theories with higher order
derivatives. In doing so, one needs to bring the Lagrangian to a first
order form, and the two most known ways to do so are either by extending
the number of the canonical momenta (the Ortogradsky formalism) or
by reducing the order of derivatives using auxiliary fields (which
we call the reduced order formalism). The first approach was considered
in \citep{Bufalo}, where the BRST quantization of the higher derivative
Podolski electrodynamics was described, using the (FJ) method to deal
with the constraints. On the other hand, recently the same model was
also considered in the reduced order formalism \citep{Reduced}, but
using the Dirac procedure for the constraints analysis. In this work,
we will work with this model also within the reduced order formalism,
but using the (FJ) formalism to work through the constraints, showing
that this approach has a very nice property, which is the clear separation
of degrees of freedom during the calculations, neatly separating the
non-massive sector from the massive (ghost) one. We believe this is
therefore the optimal approach to deal with higher derivative gauge
theories, which should also be extended to more complicated cases
such as non-Abelian theories.

This work is organized as follows. In Sec.\ref{sec2}, we review the
main conceptual aspects of Fadeev-Jackiw formalism. In Sec.\ref{sec3}
we apply the (FJ) symplectic analysis to a simple higher derivative
scalar model in both Ostrogradsky and reduced order formalisms, discussing
their equivalence at classical and quantum level, as well as the interpretation
of their degrees of freedom. In Sec.\ref{sec4} we present the reduced
order version of the (FJ) formalism in a singular higher derivative
gauge theory (Podolsky electrodynamics). Sec.\ref{sec5} contains
our conclusions and perspectives.

\section{Review of Fadeev-Jackiw formalism\label{sec2}}

We start with a brief review of the elementary aspects of the (FJ)
formalism. Starting with a Lagrangian $L(q_{i},\dot{q}_{i})$, by
means of a Lagrange transformation we define the canonical momenta
$p_{i}=\frac{\partial L}{\partial\dot{q}^{i}}$ and Hamiltonian $H=p_{i}\dot{q}^{i}-L$.
With the aim of writing the symplectic structure, the Lagrangian has
to be cast as a first order expression in the velocities $\dot{\xi}_{i}$,
where hereafter $\xi_{i}$ represents the set of all the canonical
variables in the theory (at this point, $\xi_{i}$ corresponds to
the set of the $q_{i}$ and $p_{i}$). More explicitly, the Lagrangian
has to be brought up to the form 
\begin{equation}
L\left(\xi,\dot{\xi}\right)=a_{i}(\xi)\dot{\xi}^{i}-V(\xi_{i})\thinspace,\label{eq:1}
\end{equation}
where we identify $a_{i}(\xi)=p_{i}$ and $V=H$. The equations of
motion are derived as usual from the principle of least action, 
\begin{equation}
\delta S=\int dt\left[\frac{\partial L}{\partial\xi^{i}}-\frac{d}{dt}\left(\frac{\partial L}{\partial\dot{\xi}^{i}}\right)\right]\delta\xi^{i}=0\thinspace,
\end{equation}
where $S=\int dtL$. Taking into account the explicit form of $L$
as a linear function in $\dot{\xi}^{i}$ given in \eqref{eq:1}, we
have $\frac{d}{dt}\left(\frac{\partial L}{\partial\dot{\xi}^{i}}\right)=\dot{a}_{i}\left(\xi\right)=\frac{\partial a_{i}}{\partial\xi_{j}}\dot{\xi}_{j}$
as well as $\frac{\partial L}{\partial\xi^{i}}=\frac{\partial a_{j}}{\partial\xi^{i}}\dot{\xi}^{j}-\frac{\partial V}{\partial\xi^{i}}$.
Finally, introducing the symplectic matrix $f_{ij}$, 
\begin{equation}
f_{ij}\dot{=}\frac{\partial a_{j}}{\partial{\xi}^{i}}-\frac{\partial a_{i}}{\partial{\xi}^{j}}\thinspace,
\end{equation}
we can rewrite the equations of motion as 
\begin{equation}
f_{ij}\dot{\xi}^{j}=\frac{\partial V}{\partial{\xi}^{i}}\thinspace.\label{eqmot}
\end{equation}
In the regular case, $f_{ij}$ has an inverse $f^{ij}$, and this
last equation can immediately be solved for the velocities $\dot{\xi}^{i}$
as follows, 
\begin{eqnarray}
\dot{\xi}^{i} & = & f^{ij}\frac{\partial V}{\partial{\xi}^{j}}=\{\xi^{i},V\}_{P}=\{\xi^{i},\xi^{j}\}_{P}\frac{\partial V}{\partial{\xi}^{j}}\thinspace,
\end{eqnarray}
on the other hand, when $f_{ij}$ is singular, there is no inverse
matrix since $\det[f]=0$, thus establishing the existence of zero
modes. This can be seen clearly by considering the problem of eigenvalues
and eigenvectors 
\begin{align}
 & [f]v_{a}=\omega_{a}v_{a},\\
 & \det[f-\omega_{a}I]=0,
\end{align}
from which it follows that $\det[f]=\prod_{a}\omega_{a}$. Hence if
$[f]$ is singular, $\text{det}[f]=0$ and we have null eigenvalues.
Let $\left\{ v_{n}\right\} $ be the set of linearly independent null
eigenvectors: when we multiply Eq. (\ref{eqmot}) by each of the $v_{n}$
we obtain 
\begin{equation}
v_{n}[f][\dot{\xi}]=v_{n}\frac{\partial V}{\partial[{\xi}]}=\Omega_{n}^{\left(1\right)}=0\thinspace,
\end{equation}
which represents an initial set of constraints on the dynamics. They
can be enforced by means of Lagrange multipliers $\lambda_{n}^{\left(1\right)}$,
augmenting the initial Lagrangian by the term $\sum_{n}\lambda_{n}^{\left(1\right)}\Omega_{n}^{\left(1\right)}$.
Alternatively, taking into account that the constraint does not evolve
in time ($\dot{\Omega}=0$) and that the Lagrangian is defined up
to total time derivatives, it follows that a term such as 
\begin{equation}
\frac{d\left(\lambda_{n}^{\left(1\right)}\Omega_{n}^{\left(1\right)}\right)}{dt}=\dot{\lambda}_{n}^{\left(1\right)}\Omega_{n}^{\left(1\right)}+\lambda_{n}^{\left(1\right)}\dot{\Omega}_{n}^{\left(1\right)}
\end{equation}
does not modify the dynamics, so we can actually write a first iterated
Lagrangian as 
\begin{equation}
L^{(1)}=a_{i}^{(1)}\dot{\xi}^{i}+\sum_{n}\dot{\lambda}_{n}^{\left(1\right)}\Omega_{n}^{\left(1\right)}-V^{(1)}\thinspace,
\end{equation}
where 
\begin{equation}
V^{(1)}=V|_{\Omega^{(1)}=0}\thinspace.
\end{equation}
At this point, one can enlarge the set of canonical variables $\xi_{i}$
including the $\lambda_{n}^{\left(1\right)}$. A new iteration can
be started, taking $L^{(1)}$ as the initial Lagrangian, and the procedure
continues until a non singular symplectic matrix $f_{ij}$ is obtained
-- a process which, in the case of gauge theories, involves also
the inclusion of gauge fixing conditions into the Lagrangian.

After a nonsingular symplectic matrix $f_{ij}$ is obtained at the
end of the (FJ) procedure, the transition amplitude is written as\,\citep{Toms}
\begin{equation}
Z=\int D{\xi}\sqrt{\det[f]}\exp[iS]\thinspace.\label{ampl}
\end{equation}
The crucial point to understand the previous equation is based in
the Darboux theorem, which states that by an appropriate change of
canonical coordinates ($\xi_{i}\rightarrow\xi'_{i}$), we can write
the symplectic part of the Lagrangian, in the canonical form, as 
\begin{equation}
L(\dot{Q}_{i},P_{i})=P_{i}\dot{Q}^{i}-H(Q_{i},P_{i})\thinspace,\label{eqcano}
\end{equation}
where 
\[
P_{i}\dot{Q}^{i}=\frac{1}{2}\omega^{ij}\dot{\xi}'_{i}\xi'_{j}\thinspace,
\]
$Q^{i}$ and $P_{i}$ being canonical variables obeying the standard
Poisson algebra, and $[\omega]$ the anti-symmetric block matrix,
\begin{equation}
[\omega]=\left[\begin{array}{cc}
0 & -I\\
I & 0
\end{array}\right].
\end{equation}
In fact, Eq. (\ref{eqcano}) can be written in the following form,
\begin{equation}
L(\xi')=a'^{i}\dot{\xi}'_{i}-H(\xi')\thinspace,
\end{equation}
where 
\begin{equation}
a'^{i}=\frac{1}{2}\xi'_{j}\omega^{ji}\thinspace.
\end{equation}
Here we can identify the arbitrary vector potential (one-form) as
\begin{equation}
a'=a'^{i}d\xi'_{i},
\end{equation}
whose associated field strength (two-form) is given by 
\begin{eqnarray}
da' & = & \frac{\partial a'^{i}}{\partial\xi'_{j}}d\xi'_{i}d\xi'_{j}\\
 & = & \frac{1}{2}\left(\frac{\partial\xi'_{i}}{\partial\xi_{a}}\omega^{ij}\frac{\partial\xi'_{j}}{\partial\xi_{b}}\right)d\xi_{a}d\xi_{b}\thinspace.
\end{eqnarray}
The fact that the action $S=\int a_{i}d\xi^{i}-\int Hdt$ is invariant
under canonical transformations leads us to define the symplectic
matrix as 
\begin{equation}
f_{ij}\dot{=}\frac{\partial\xi'_{i}}{\partial\xi_{a}}\omega_{ij}\frac{\partial\xi'_{j}}{\partial\xi_{b}}\thinspace,
\end{equation}
while, by the Schwinger variational principle of quantum action, $\delta Z=\langle\delta\hat{S}\rangle=\delta SZ$,
we have 
\begin{equation}
Z=\int DQDP\exp\biggl[i\int dt\biggr(P_{i}\dot{Q}^{i}-H(Q_{i},P_{i})\biggl)\biggr]\thinspace,
\end{equation}
or, in other words, $Z=\int D{\xi'}\exp[iS']$, where $S'=\int a'_{i}d\xi'^{i}-\int H'dt$.
Therefore, by a canonical transformation ($\xi'_{i}\rightarrow\xi_{i}$),
we write 
\begin{equation}
Z=\int D{\xi}\det\left(\frac{\partial\xi'^{i}}{\partial\xi_{j}}\right)\exp[iS]\thinspace,
\end{equation}
wherein we see that 
\begin{equation}
\det\left(\frac{\partial\xi'^{i}}{\partial\xi_{j}}\right)=\sqrt{\det\left(\frac{\partial\xi'_{i}}{\partial\xi_{a}}\omega_{ij}\frac{\partial\xi'_{j}}{\partial\xi_{b}}\right)}=\sqrt{\det[f]}.
\end{equation}
As stated in\,\citep{Toms}, it is important that the final result
actually does not depend on the explicit form of the transformation
($\xi'_{i}\rightarrow\xi_{i}$), but only on the symplectic structure
of the theory, which is solved by the (FJ) procedure.

\section{Toy model as a prof of concept\label{sec3}}

In this section, we consider a rather simple higher derivative theory,
based on a massless real scalar field. The aim is to gain insight
in the physical interpretations of such theories, and to present in
a simpler setting the procedure to be considered in connection to
the Podolsky electrodynamics in the next section. We will work out
both the Ostrogradsky and the reduced order approach, and we will
explicitly verify that both routes lead to the same quantum theory.
In the literature, the first order form of higher derivative theories
was explored in \citep{Krug}. The connections between the Ostrogradsky
formalism (starting with a fourth order Lagrangian), the reduction
of order formalism with an auxiliary field (starting with a first
order Lagrangian, directly suitable to the application of the (FJ)
method), and the final first order description (Hamiltonian), should
be such that in any description we have the same propagating degrees
of freedom, which in the present case are two: one being the original
massless and the other one, massive, whose physical interpretation
is of a ghost (unphysical) mode. We will also briefly comment on some
recent ideas on how to interpret the presence of this ghost mode.

\subsection{Ostrogradsky formalism}

We being with the Lagrangian density ${\cal L}_{Ostro}$, 
\begin{equation}
{\cal L}_{Ostro}=\frac{1}{2}\partial_{\mu}\phi\left(1+\frac{\Box}{m^{2}}\right)\partial^{\mu}\phi\thinspace,
\end{equation}
so the corresponding, fourth order equation of motion is given by
\begin{equation}
\Box(\Box+m^{2})\phi=0\thinspace.\label{eqmotion}
\end{equation}
According to the Noether theorem, the conserved quantity corresponding
to the time translation invariance of the action is the Hamiltonian
density 
\begin{equation}
{\cal H}_{Ostro}=\pi\partial_{0}\phi+P\partial_{0}Q-{\cal L}_{Ostro}\thinspace,\label{eq:HOstro}
\end{equation}
where the additional canonical coordinate $Q=\partial_{0}\phi$ was
introduced to account for the higher order time derivatives. The canonical
momenta are given by 
\begin{equation}
\pi=\left(1+\frac{\Box}{m^{2}}\right)\partial_{0}\phi,\thinspace\thinspace P=-\frac{\Box}{m^{2}}\phi\thinspace,
\end{equation}
and therefore 
\begin{equation}
{\cal H}_{Ostro}(\phi,Q;\pi,P)=\pi Q-\frac{1}{2}m^{2}P^{2}-P\partial_{k}\partial^{k}-\frac{1}{2}D^{2}-\partial_{k}\phi\partial^{k}\phi\thinspace.
\end{equation}
The first order Lagrangian can be written as 
\begin{equation}
{\cal L}_{Ostro}=\pi\partial_{0}\phi+P\partial_{0}Q-{\cal H}_{Ostro}\thinspace,
\end{equation}
where the canonical one form of the symplectic variables $\xi=(\phi,\pi,Q,P)$
corresponds to 
\begin{equation}
a_{\phi}=\pi,\quad a_{\pi}=0,\quad a_{Q}=P,\quad a_{P}=0\thinspace.
\end{equation}
Therefore, we obtain the symplectic matrix $f_{ij}=\frac{\partial a_{j}}{\partial{\xi}^{i}}-\frac{\partial a_{i}}{\partial{\xi}^{j}}$,
\begin{equation}
[f]=\left[\begin{array}{c|cccc}
 & \phi & \pi & Q & P\\
\hline \phi & 0 & -1 & 0 & 0\\
\pi & 1 & 0 & 0 & 0\\
Q & 0 & 0 & 0 & -1\\
P & 0 & 0 & 1 & 0
\end{array}\right]\delta^{3}(\vec{x}-\vec{y})\thinspace,
\end{equation}
and, as $\det[f]=1$, the inverse matrix exists, and can be readily
obtained as $[f]^{-1}=-[f]$. As a consequence, the fundamental non
null Poisson brackets read 
\begin{equation}
\{\phi(x),\pi(y)\}_{P}=\delta^{3}(\vec{x}-\vec{y}),\quad\{Q(x),P(y)\}_{P}=\delta^{3}(\vec{x}-\vec{y}).
\end{equation}

Now, going to the quantum language, the transition amplitude is given
in view of Eq. (\ref{ampl}), as 
\begin{align}
Z_{Ostro}= & \int D\phi D\pi DQDP\exp\big\{ i\int d^{4}x[\pi\partial_{0}\phi+P\partial_{0}Q-{\cal H}_{Ostro}]\big\}\nonumber \\
= & \int D\phi D\pi DQDP\exp\big\{ i\int d^{4}x\thinspace\left[\pi\partial_{0}\phi+P\partial_{0}Q-\pi Q+\frac{1}{2}m^{2}P^{2}+P\partial_{k}\partial^{k}\phi+\right.\nonumber \\
 & +\left.+\frac{1}{2}Q^{2}\vphantom{\frac{1}{2}}\partial_{k}\phi\partial^{k}\phi\right]\big\}.
\end{align}
After integration in $DQD\pi$ , and completing the squares we obtain
as our final result the gaussian functional, 
\begin{align}
Z_{Ostro}= & \int D\phi DP\exp\left\{ i\int d^{4}x\thinspace\left[\frac{1}{2}m^{2}\left(P+\frac{\partial_{\mu}\partial^{\mu}\phi}{m^{2}}\right)^{2}-\left(\frac{\Box}{m}\phi\right)^{2}-\frac{1}{2}\phi\Box\phi\right]\right\} \\
=\thinspace & N\int D\phi\exp\left\{ -i\int d^{4}x\thinspace\phi\Box\left(1+\frac{\Box}{m^{2}}\right)\phi\right\} \\
= & \thinspace N\det\left[-\frac{1}{16}\Box(\Box+m^{2})\right]\thinspace.\label{eqOstro}
\end{align}

\subsection{Reduced order with an auxiliary field}

Instead of dealing with the higher derivatives via the Ostrograsdky
method, one may also introduce an auxiliary field $Z$, starting with
the Lagrangian 
\begin{equation}
{\cal L}_{red}=\frac{1}{2}\phi\Box Z-\frac{1}{8}m^{2}\phi\phi+\frac{1}{4}m^{2}\phi Z-\frac{1}{8}m^{2}ZZ\thinspace,\label{reducedL}
\end{equation}
whose corresponding equations of motion are given by 
\begin{equation}
\left(1+2\frac{\Box}{m^{2}}\right)\phi=Z,\thinspace\thinspace\left(1+2\frac{\Box}{m^{2}}\right)Z=\phi\thinspace.\label{eqaclo}
\end{equation}
These set of coupled equations are equivalent to Eq. (\ref{eqmotion}),
as can be seen by direct substitution. The canonical Hamiltonian is
given by ${\cal H}_{red}=\pi\partial_{0}\phi+\theta\partial_{0}Z-{\cal L}_{red}$
with the respective canonical momenta defined as 
\begin{align}
 & \pi\dot{=}\frac{\partial{\cal L}_{red}}{\partial(\partial_{0}\phi)}=-\frac{1}{2}\partial_{0}Z\thinspace,\\
 & \theta\dot{=}\frac{\partial{\cal L}_{red}}{\partial(\partial_{0}Z)}=-\frac{1}{2}\partial_{0}\phi\thinspace,
\end{align}
or, more explicitly, 
\begin{equation}
{\cal H}_{red}=-2\pi\theta+\frac{1}{2}\partial_{i}\phi\partial^{i}Z+\frac{1}{8}m^{2}\phi\phi-\frac{1}{4}m^{2}\phi Z+\frac{1}{8}m^{2}ZZ\thinspace.
\end{equation}
Therefore the canonical one form of the symplectic variables $\xi=(\phi,\pi,Z,\theta)$
is given by 
\begin{equation}
a_{\phi}=\pi,\quad a_{\pi}=0,\quad a_{Z}=\theta,\quad a_{\theta}=0\thinspace,
\end{equation}
and the corresponding symplectic matrix is 
\begin{equation}
[f]=\left[\begin{array}{c|cccc}
 & \phi & \pi & Z & \theta\\
\hline \phi & 0 & -1 & 0 & 0\\
\pi & 1 & 0 & 0 & 0\\
Z & 0 & 0 & 0 & -1\\
\theta & 0 & 0 & 1 & 0
\end{array}\right]\delta^{3}(\vec{x}-\vec{y})\thinspace,
\end{equation}
which again is a non-singular, unitary determinant matrix, with inverse
$[f]^{-1}=-[f]$. The corresponding fundamental non null Poisson brackets
are 
\begin{equation}
\{\phi(x),\pi(y)\}_{P}=\delta^{3}(\vec{x}-\vec{y}),\quad\{Z(x),\theta(y)\}_{P}=\delta^{3}(\vec{x}-\vec{y}).
\end{equation}

Quantization is achieved by calculating the transition amplitude which
in this case reads 
\begin{align}
Z_{red}= & \int D\phi D\pi DZD\theta\exp\left\{ i\int d^{4}x\left[\pi\partial_{0}\phi+\theta\partial_{0}Z-{\cal H}_{red}\right]\right\} \\
= & \int D\phi D\pi DZD\theta\exp\left\{ i\int d^{4}x\left[\pi\partial_{0}\phi+\theta\partial_{0}Z+2\pi\theta-\frac{1}{2}\partial_{i}\phi\partial^{i}Z-\frac{1}{8}m^{2}\phi\phi\right.\right.\nonumber \\
 & \left.\left.+\frac{1}{4}m^{2}\phi Z-\frac{1}{8}m^{2}ZZ\right]\right\} \thinspace.
\end{align}
Integrating in $D\pi D\theta$, one obtains 
\[
Z_{red}=\int D\phi DZ\exp\left\{ i\int d^{4}x\left[-\frac{1}{2}\partial_{\mu}\phi\partial^{\mu}Z-\frac{1}{8}m^{2}\phi\phi+\frac{1}{4}m^{2}\phi Z-\frac{1}{8}m^{2}ZZ\right]\right\} \thinspace,
\]
and therefore 
\begin{align}
Z_{red} & =\int D\phi DZ\exp\left\{ i\int d^{4}x\left[\begin{array}{cc}
\phi & Z\end{array}\right]\left[\begin{array}{cc}
-\frac{m^{2}}{8} & (\frac{\Box}{4}+\frac{m^{2}}{8})\\
(\frac{\Box}{4}+\frac{m^{2}}{8}) & -\frac{m^{2}}{8}
\end{array}\right]\left[\begin{array}{c}
\phi\\
Z
\end{array}\right]\right\} \thinspace,\label{eqred}\\
 & =N\int D\phi\exp\left\{ -i\int d^{4}x\phi\Box\left(1+\frac{\Box}{m^{2}}\right)\phi\right\} \thinspace,
\end{align}
which reduces to the determinant of the square matrix appearing in
Eq.\,\eqref{eqred}. The determinant of course involves both the
discrete matrix indices as well as the continuous spacetime indices
(coordinates): calculating explicitly the first part gives 
\begin{equation}
\det\left[\begin{array}{cc}
-\frac{m^{2}}{8} & (\frac{\Box}{4}+\frac{m^{2}}{8})\\
(\frac{\Box}{4}+\frac{m^{2}}{8}) & -\frac{m^{2}}{8}
\end{array}\right]=\det\left[-\frac{1}{16}\Box(\Box+m^{2})\right]\thinspace,
\end{equation}
which agrees with Eq. (\ref{eqOstro}). We therefore verity that the
equivalence between the classical equations of motion in the Ostrogradsky
and Reduction of order prescriptions, seen in Eqs. (\ref{eqmot})
and (\ref{eqaclo}), hold also at the quantum level, when we compare
the transition amplitude obtained in both prescriptions.

\subsection{Characterization of the degrees of freedom}

It is a common feature of higher derivatives theories to present additional,
non physical degrees of freedom. This can be clearly seen in the present
model. We choose to use the reduced order formalism as discussed in
the previous subsection. The coupled equations of motion for the $\phi$
and $Z$ field, given in Eq.\,(\ref{eqaclo}), can be written in
matrix notation as 
\begin{align}
M\left(\begin{array}{c}
\phi\\
Z
\end{array}\right) & =\left(\begin{array}{c}
0\\
0
\end{array}\right),
\end{align}
where 
\begin{align}
M & =\left(\begin{array}{cc}
\left(1+\frac{2}{m^{2}}\square\right) & -1\\
-1 & \left(1+\frac{2}{m^{2}}\square\right)
\end{array}\right).
\end{align}
The dynamics can be rewritten in order to make manifest the fact that
it involves two independent degrees of freedom. At the matrix level,
this amounts to the problem of diagonalization of the matrix $M$.
The eigenvalues of $M$ are determined by the equation $\det[M-\lambda I]=0$,
whose solutions are 
\begin{align}
\lambda_{\pm}= & \left(1+\frac{2}{m^{2}}\square\right)\mp1\thinspace.
\end{align}
So the matrix $M$ is unitarily equivalent to a matrix describing
two degrees of freedom, one being massless, and the other massive.
Indeed, by means of a linear transformation, 
\begin{align}
 & \phi=\alpha A+\beta B\thinspace,\\
 & Z=\alpha A-\beta B\thinspace,
\end{align}
the Lagrangian in Eq. (\ref{reducedL}) can be brought to the following
form, 
\begin{equation}
\mathcal{L}_{red}^{\prime}=\alpha^{2}\left[\frac{1}{2}A\square A\right]-\beta^{2}\left[\frac{1}{2}B\square B+\frac{{m_{p}}^{2}}{2}B^{2}\right]\thinspace.
\end{equation}
This last equation explicitly separates the two degrees of freedom
present in the model. For real $\alpha$ and $\beta$, clearly the
$B$ mode appears with a ``wrong sign'' in the Lagrangian, and will
in fact violate the stability of the Hamiltonian. Therefore, $B$
should be interpreted as a non physical (ghost) degree of freedom.

The presence of ghosts is a longstanding issue in the quantization
of higher derivative models. Recently, it has been pointed out that,
at least in the free case, these ghosts could be reinterpreted as
physical particles after a proper complexification: this was discussed
for the Pais-Uhlenbeck oscillator in\,\citep{Complex}. In the present
case, one may note that the choice 
\begin{align}
\phi & =A+iB\thinspace,\nonumber \\
Z & =A-iB\thinspace,
\end{align}
recovers the stability of the Hamiltonian.

If we try to interpreted the imaginary part of the field $\phi$ as
a massive physical degree of freedom, so that both $A$ and $B$ are
real degrees of freedom associated with the real and imaginary parts
of the field $\phi$, it may seem that by complexifying the original
Lagrangian we are increasing the degrees of freedom to four (complex
$\phi$ and $Z$ fields). Actually, the balance in the degrees of
freedom can be preserved with the introduction the condition $Z=\bar{\phi}$
by means of a Lagrange multiplier $\lambda$ into the reduced order
complex scalar Lagrangian, 
\begin{align}
\mathcal{L} & =-\frac{1}{2}\phi\square Z+\frac{1}{8}m^{2}\phi\phi-\frac{1}{4}m^{2}\phi Z+\frac{1}{8}m^{2}ZZ+\lambda\left(Z-\bar{\phi}\right),\label{complex_lagrangian_Z}
\end{align}
$\phi$ and $Z$ being now complex fields. The equations of motion
are given by 
\begin{align}
 & \left(1+\frac{2}{m^{2}}\square\right)\phi=Z-\frac{4}{m^{2}}\lambda\thinspace,\\
 & \left(1+\frac{2}{m^{2}}\square\right)Z=\phi\thinspace,\\
 & Z=\bar{\phi}\thinspace,
\end{align}
which can be combined and brought into the form 
\begin{align}
 & \square\left(1+\frac{1}{m^{2}}\square\right)\phi=0\thinspace,\\
 & \square\left(\phi+Z\right)=0\thinspace,\\
 & \left(\square+m^{2}\right)\left(\phi-Z\right)=0\thinspace,
\end{align}
where we conclude that $\lambda=0$, $\phi=A+iB$ and $Z=A-iB$. Substituting
this in (\ref{complex_lagrangian_Z}), we end up with 
\begin{align}
\mathcal{L}=\frac{1}{2}A\square A+\frac{1}{2}B\square B+\frac{1}{2}m^{2}B^{2}.
\end{align}
In summary: as $\phi$ and $Z$ are complex fields we start with four
degrees of freedom described by the complex Lagrangian (\ref{complex_lagrangian_Z}),
while the higher derivative real scalar theory has only two degrees
of freedom. We match the number of degrees of freedom in both formulation
by enforcing the condition $Z=\bar{\phi}$ via a Lagrange multiplier.

A more general prescription to quantize higher derivative theories,
circumventing the problem of the stability of the Hamiltonian, have
been discussed in\,\citep{Russos}, using the concept of Lagrangian
anchors. Essentially, it involves an extension of the Noether theorems,
defining a class of conserved quantities associated with a given symmetry.
For time translations, this procedure can lead to different conserved
quantities which could be in principle be identified with a Hamiltonian,
some of them would have regularizing properties but will break the
stability because the energy is not bounded from below, whereas the
other recovers the stability but loses the regularizing properties,
seen in the self-energy of a particles and ultraviolet divergences.
It would be an interesting endeavor to investigate this approach for
more involved models, something that we will not try in this work.

\section{HD Podolsky theory in the (FJ) formalism\label{sec4}}

Although the (FJ) formalism does not implement major changes in the
quantization process of a regular theory, in a singular theory there
might be considerable simplifications when adopting the symplectic
formalism instead of the usual (DB) algorithm. We apply the (FJ) method
to discuss the quantization of the Podolsky electrodynamics \citep{Gaugefixing}
but, differently from what was done in\,\citep{Bufalo}, we start
by writing the theory in the reduction of order formalism, by means
of the introduction of an additional auxiliary field $B^{\mu}$, following\,\citep{Reduced}.
We will show that this technique allows us to write the sympletic
matrix in a block structure, thus clearly separating the Maxwell and
Proca sectors. This makes the treatment of the different degrees of
freedom of the model particularly simple and clear.

Concretely, we start with, 
\begin{equation}
{\cal L}_{red}=-\frac{1}{4}F^{\mu\nu}F_{\mu\nu}-\frac{a^{2}}{2}B_{\mu}B^{\mu}+a^{2}\partial_{\mu}B_{\nu}F^{\mu\nu}\thinspace,\label{Ost1}
\end{equation}
where 
\begin{equation}
F_{\mu\nu}=\partial_{\mu}A_{\nu}-\partial_{\nu}A_{\mu}\thinspace.
\end{equation}
Up to surface terms, we can also write 
\begin{equation}
{\cal L}_{red}=\frac{1}{2}A^{\mu}(\eta_{\mu\nu}\Box-\partial_{\mu}\partial_{\nu})A^{\nu}-\frac{a^{2}}{2}B_{\mu}B^{\mu}-a^{2}B^{\mu}(\eta_{\mu\nu}\Box-\partial_{\mu}\partial_{\nu})A^{\mu}\thinspace,
\end{equation}
which leads directly to the coupled equations of motion 
\begin{equation}
(\eta_{\mu\nu}\Box-\partial_{\mu}\partial_{\nu})A^{\nu}=a^{2}(\eta_{\mu\nu}\Box-\partial_{\mu}\partial_{\nu})B^{\nu}\thinspace,\label{coupleeq}
\end{equation}
and 
\begin{equation}
(\eta_{\mu\nu}\Box-\partial_{\mu}\partial_{\nu})A^{\nu}=-B_{\mu}\thinspace.\label{coupleeq1}
\end{equation}
A direct consequence of the last equation is that $\partial_{\mu}B^{\mu}=0$.
Additionally, one may decouple the previous two equations, obtaining
\begin{equation}
(1+a^{2}\Box)(\eta_{\mu\nu}\Box-\partial_{\mu}\partial_{\nu})A^{\nu}=0;\thinspace\thinspace(1+a^{2}\Box)B_{\mu}=0\thinspace.\label{4order}
\end{equation}
Classically the reduced order Lagrangian density ${\cal L}_{red}$
is equivalent to the following Ostrogradsky Lagrangian density, up
to surface terms, 
\begin{equation}
{\cal L}_{Ostro}=-\frac{1}{4}F^{\mu\nu}(1+a^{2}\Box)F_{\mu\nu}=-\frac{1}{4}F^{\mu\nu}F_{\mu\nu}+\frac{a^{2}}{2}\partial_{\nu}F^{\mu\nu}\partial^{\rho}F_{\mu\rho}\thinspace.\label{LagOstro}
\end{equation}
Also, the classical coupled equations of motion (\ref{coupleeq})
and (\ref{coupleeq1}) can be written as 
\begin{equation}
\left[\begin{array}{cc}
T_{\mu\nu} & a^{2}T_{\mu}^{\nu}\\
T_{\nu}^{\mu} & -\eta^{\mu\nu}
\end{array}\right]\left[\begin{array}{c}
A^{\nu}\\
B_{\nu}
\end{array}\right]=0,
\end{equation}
wherein we have the definition $T_{\mu\nu}\dot{=}\eta_{\mu\nu}\Box-\partial_{\nu}\partial_{\nu}$.
Implicitly in the analysis we have a problem of eigenvalues and eigenvectors
and the diagonalization of a matrix since 
\begin{equation}
\det\left[\begin{array}{cc}
T_{\mu\nu} & a^{2}T_{\mu}^{\nu}\\
T_{\nu}^{\mu} & -\eta^{\mu\nu}
\end{array}\right]=\det\left[\begin{array}{cc}
T_{\mu\nu} & 0\\
0 & -(T^{\mu\nu}+a^{2}\eta^{\mu\nu})
\end{array}\right]=-3\Box(1+a^{2}\Box),
\end{equation}
making explicit the Maxwell (the $T_{\mu\nu}$ factor) and Proca (the
$-(T^{\mu\nu}+a^{2}\eta^{\mu\nu})$ factor) physical degrees of freedom
(2+3, respectively) of the theory, as well as the problem of instability
due to the negative sign of the massive mode.

Due to the fact that ${\cal L}_{red}$ is of second order, we can
define the usual canonical momenta 
\begin{equation}
\pi^{i}=\frac{\partial{\cal L}_{red}}{\partial\dot{A}_{i}}=F^{0i}+a^{2}(\partial^{0}B^{i}-\partial^{i}B^{0}),\quad\theta^{i}=\frac{\partial{\cal L}_{red}}{\partial\dot{B}_{i}}=a^{2}F^{0i},
\end{equation}
leading to 
\[
{\cal L}_{red}=\frac{1}{a^{2}}\pi^{i}\theta_{i}-\frac{1}{2a^{4}}\theta^{i}\theta_{i}-\frac{1}{4}F^{ij}F_{ij}+a^{2}\partial_{i}B_{j}F^{ij}-\frac{a^{2}}{2}B_{\mu}B^{\mu}\thinspace.
\]
By a Legendre transform, we obtain the canonical Hamiltonian 
\begin{equation}
{\cal H}_{red}=\pi_{i}\dot{A}^{i}+\theta_{i}\dot{B}^{i}-{\cal L}_{red}\thinspace,
\end{equation}
which, up to surface terms, leads to 
\begin{equation}
{\cal H}_{red}(A_{i},\pi_{i},B_{i},\theta_{i},A_{0},B_{0})=\frac{1}{a^{2}}\pi^{i}\theta_{i}+\frac{1}{2a^{4}}\theta^{i}\theta_{i}+\frac{1}{4}F^{ij}F_{ij}-a^{2}\partial_{i}B_{j}F^{ij}+\frac{a^{2}}{2}B_{\mu}B^{\mu}-A^{0}\partial^{i}\pi_{i}-B^{0}\partial^{i}\theta_{i}\thinspace.
\end{equation}

We can now construct the symplectic structure in the (FJ) formalism,
starting by writing 
\begin{equation}
{\cal L}_{red}=\frac{1}{a^{2}}\pi_{i}\dot{A}^{i}+\theta_{i}\dot{B}^{i}-{\cal V}^{(0)}\thinspace,
\end{equation}
where 
\begin{equation}
{\cal V}^{(0)}=\pi^{i}\theta_{i}+\frac{1}{2a^{4}}\theta^{i}\theta_{i}+\frac{1}{4}F^{ij}F_{ij}-a^{2}\partial_{i}B_{j}F^{ij}+\frac{a^{2}}{2}B_{\mu}B^{\mu}-A^{0}\partial^{i}\pi_{i}-B^{0}\partial^{i}\theta_{i}\thinspace.
\end{equation}
The symplectic variables are up to this point $\xi=(A_{i},\pi_{i},B_{i},\theta_{i},A_{0},B_{0})$
and the canonical one-form is given by 
\begin{equation}
a_{A_{i}}=\pi_{i},\quad a_{\pi_{i}}=0,\quad a_{B_{i}}=\theta_{i},\quad a_{\theta_{i}}=0,\quad a_{A_{0}}=0,\quad a_{B_{0}}=0\thinspace,
\end{equation}
therefore, the symplectic matrix can be written as 
\begin{equation}
[f]=\left[\begin{array}{c|cccccc}
 & A_{j} & \pi_{j} & B_{j} & \theta_{j} & A_{0} & B_{0}\\
\hline A_{i} & 0 & -\delta_{ij} & 0 & 0 & 0 & 0\\
\pi_{i} & \delta_{ij} & 0 & 0 & 0 & 0 & 0\\
B_{i} & 0 & 0 & 0 & -\delta_{ij} & 0 & 0\\
\theta_{i} & 0 & 0 & \delta_{ij} & 0 & 0 & 0\\
A_{0} & 0 & 0 & 0 & 0 & 0 & 0\\
B_{0} & 0 & 0 & 0 & 0 & 0 & 0
\end{array}\right]\delta^{3}(\vec{x}-\vec{y}).
\end{equation}
Clearly, $\det[f]=0$ signaling a singular system, as expected. The
following eigenvectors have null eigenvalues, 
\begin{align}
u= & ({\bold0},{\bold0},{\bold0},{\bold0},u^{13},0)\thinspace,\\
v= & ({\bold0},{\bold0},{\bold0},{\bold0},0,v^{14})\thinspace,
\end{align}
and the respective constraint equations are 
\begin{subequations}
\label{eq:2ndOrderConst}
\begin{align}
\Omega_{1}= & \int dxdy\thinspace u\frac{\delta{\cal V}^{(0)}(y)}{\delta A_{0}(x)}=\int dx\thinspace u^{13}\partial^{i}\pi_{i}=0\thinspace,\\
\Omega_{2}= & \int dxdy\thinspace v\frac{\delta{\cal V}^{(0)}(y)}{\delta B_{0}(x)}=\int dx\thinspace u^{13}(a^{2}B_{0}-\partial^{i}\theta_{i})=0\thinspace.
\end{align}
\end{subequations}
We enforce the previous constraint equations into ${\cal L}_{red}$
using Lagrange multipliers, 
\begin{equation}
{\cal L}_{red}=\pi_{i}\dot{A}^{i}+\theta_{i}\dot{B}^{i}+\dot{\lambda}^{a}\Omega_{a}-{\cal V}^{(2)},\quad a=1,2\thinspace,
\end{equation}
where 
\begin{equation}
{\cal V}^{(2)}={\cal V}^{(0)}|_{\Omega_{a}=0}=\frac{1}{a^{2}}\pi^{i}\theta_{i}+\frac{1}{2a^{4}}\theta^{i}\theta_{i}+\frac{1}{4}F^{ij}F_{ij}-a^{2}\partial_{i}B_{j}F^{ij}-\frac{a^{2}}{2}B_{0}B^{0}+\frac{a^{2}}{2}B_{i}B^{i}\thinspace.
\end{equation}
So from this augmented symplectic structure, we have the following
one form vectors 
\begin{equation}
a_{A_{i}}^{(2)}=\pi_{i},\quad a_{\pi_{i}}^{(2)}=0,\quad a_{B_{0}}^{(2)}=0,\quad a_{B_{i}}^{(2)}=\theta_{i},\quad a_{\theta_{i}}^{(2)}=0,\quad a_{\lambda_{1}}^{(2)}=\partial^{i}\pi_{i},\quad a_{\lambda_{2}}^{(2)}=(a^{2}B_{0}-\partial^{i}\theta_{i})\thinspace.
\end{equation}
At this point, when calculating the symplectic matrix, one realizes
the main advantage in the present formalism, since $\left[f\right]$
turns out to be a block diagonal matrix 
\begin{equation}
[f]=\left[\begin{array}{cc}
[M] & {\bold0}\\
{\bold0} & [P]
\end{array}\right]\thinspace,\label{eq:blockdiagonal}
\end{equation}
where $\left[M\right]$ corresponds to the massless Maxwell sector
of the theory, 
\begin{equation}
[M]=\left[\begin{array}{c|ccc}
 & A_{j} & \pi_{j} & \lambda_{1}\\
\hline A_{i} & 0 & -\delta_{ij} & \partial_{i}\\
\pi_{i} & \delta_{ij} & 0 & 0\\
\lambda_{1} & \partial_{j} & 0 & 0
\end{array}\right]\delta^{3}(\vec{x}-\vec{y})\thinspace,
\end{equation}
and $\left[P\right]$ to the massive Proca sector 
\begin{equation}
[P]=\left[\begin{array}{c|cccc}
 & B_{j} & \theta_{j} & B_{0} & \lambda_{2}\\
\hline B_{i} & 0 & -\delta_{ij} & 0 & \partial_{i}\\
\theta_{i} & \delta_{ij} & 0 & 0 & 0\\
B_{0} & 0 & 0 & 0 & -a^{2}\\
\lambda_{2} & \partial_{j} & 0 & a^{2} & 0
\end{array}\right]\delta^{3}(\vec{x}-\vec{y})\thinspace.\label{eq:matrixP}
\end{equation}
Needless to say, the structure of $\left[f\right]$ implies that 
\begin{equation}
\det[f]=\det[M]\det[P]\thinspace.
\end{equation}
The neat separation between the Maxwell and Proca sectors is a distinctive
feature of the (FJ) formalism applied to the reduced order Podolsky
electrodynamics, which does not happen within the Ortogradsky formalism\,\citep{Bufalo}.

First, let us work with the Maxwell sector. As expected, $\det[M]=0$
so $[M]$ is singular, and the null eigenvector is of the form $v=(0,v_{j}^{\pi},v^{\lambda_{1}})$,
$j=1,2,3$, corresponding to the constraint equation 
\begin{equation}
\int dxdy\thinspace v_{i}^{\pi}\frac{\delta{\cal V}^{(2)}(y)}{\delta A_{i}(x)}=\int dx\thinspace\partial_{i}\partial_{j}v^{\lambda_{1}}\left[-\partial_{i}F^{ij}+\frac{a^{2}}{2}\partial_{i}\left(\partial^{i}B^{j}-\partial^{j}B^{j}\right)\right]=0\thinspace.
\end{equation}
This zero mode does not generate any additional constraints and, consequently,
the symplectic matrix remains singular, which is a characteristic
of gauge theories: a gauge fixing condition should be introduced in
order to obtain a non singular symplectic matrix. Inspired by the
form of the fourth-order equations of motion for $A_{\mu}$, Eq.\,(\ref{4order}),
as well as the analysis presented in\,\citep{Bufalo}, we use generalized
Coulomb gauge fixing conditions in the form 
\begin{equation}
A_{0}=0,\quad\Omega_{3}=(1+a^{2}\Box)\vec{\nabla}\vec{A}=0\thinspace.\label{eq:GaugeFixing}
\end{equation}
For more details on the gauge fixing of the Podolsky theory we refer
the reader to\,\citep{Gaugefixing}. When this gauge condition is
included in ${\cal L}_{red}$ using a Lagrange multiplier $\lambda_{3}\Omega_{3}$,
we obtain the following $[M]$ matrix for the Maxwell sector 
\begin{equation}
[M]=\left[\begin{array}{c|cccc}
 & A_{j} & \pi_{j} & \lambda_{3} & \lambda_{2}\\
\hline A_{i} & 0 & -\delta_{ij} & 0 & \partial_{i}\\
\pi_{i} & \delta_{ij} & 0 & (1+a^{2}\vec{\nabla}^{2})\partial_{i} & 0\\
\lambda_{3} & 0 & (1+a^{2}\vec{\nabla}^{2})\partial_{j} & 0 & 0\\
\lambda_{2} & \partial_{j} & 0 & 0 & 0
\end{array}\right]\delta^{3}(\vec{x}-\vec{y})\thinspace,
\end{equation}
which is a regular matrix, with $\det[M]=\left[(1+a^{2}\vec{\nabla}^{2})\vec{\nabla}^{2}\right]^{2}$,
and its inverse can be calculated almost immediately 
\begin{multline}
[M]^{-1}=\frac{1}{(1+a^{2}\vec{\nabla}^{2})\vec{\nabla}^{2}}\times\\
\times\left[\begin{array}{c|cccc}
 & A_{j} & \pi_{j} & \lambda_{3} & \lambda_{2}\\
\hline A_{i} & 0 & -(1+a^{2}\vec{\nabla}^{2})\vec{\nabla}^{2}\delta_{ij}+\partial_{i}\partial_{j} & 0 & \partial_{i}\\
\pi_{i} & (1+a^{2}\vec{\nabla}^{2})\vec{\nabla}^{2}\delta_{ij}-\partial_{i}\partial_{j} & 0 & \partial_{i} & 0\\
\lambda_{3} & 0 & \partial_{j} & 0 & 1\\
\lambda_{2} & \partial_{j} & 0 & 1 & 0
\end{array}\right]\delta^{3}(\vec{x}-\vec{y})\thinspace.
\end{multline}
From this, one easily identifies the Dirac brackets between the dynamics
variables in the generalized Lorenz gauge 
\begin{equation}
\{A_{i},\pi_{j}\}_{D}=\left[-\delta_{ij}+\frac{\partial_{i}\partial_{j}}{(1+a^{2}\vec{\nabla}^{2})\vec{\nabla}^{2}}\right]\delta^{3}(\vec{x}-\vec{y}).
\end{equation}

Now, we consider the Proca sector. One way to calculate the determinant
of $\left[P\right]$ is to notice that, for any block matrix of the
form 
\begin{equation}
P_{n\times n}=\left(\begin{array}{cc}
A_{m\times m} & B_{m\times n-m}\\
C_{n-m\times m} & D_{m\times m}
\end{array}\right),
\end{equation}
if $D$ has an inverse, the following identity holds 
\begin{equation}
\left(\begin{array}{cc}
A & B\\
C & D
\end{array}\right)\times\left(\begin{array}{cc}
I & 0\\
-D^{-1}C & I
\end{array}\right)=\left(\begin{array}{cc}
A-BD^{-1}C & B\\
0 & D
\end{array}\right)\thinspace,
\end{equation}
and therefore 
\begin{equation}
\det P=\det\left(A-BD^{-1}C\right)\det D\thinspace.
\end{equation}
Applied to Eq.\,\eqref{eq:matrixP}, this leads to $\det{[P]}=a^{4}$.
The Proca sector is therefore regular, and we obtain the following
inverse symplectic matrix 
\begin{equation}
[P]^{-1}=\left[\begin{array}{c|cccc}
 & B_{j} & \theta_{j} & B_{0} & \lambda_{2}\\
\hline B_{i} & 0 & -\delta_{ij} & -\frac{1}{a^{2}}\partial_{i} & 0\\
\theta_{i} & \delta_{ij} & 0 & 0 & 0\\
B_{0} & \frac{1}{a^{2}}\partial_{j} & 0 & 0 & \frac{1}{a^{2}}\\
\lambda_{2} & 0 & 0 & -\frac{1}{a^{2}} & 0
\end{array}\right]\delta^{3}(\vec{x}-\vec{y})\thinspace,
\end{equation}
corresponding to the following Dirac brackets between the dynamics
variables, 
\begin{equation}
\{B_{i},\theta_{j}\}_{D}=-\delta_{ij}\delta^{3}(\vec{x}-\vec{y})\thinspace.
\end{equation}

From now on we are ready to construct the quantum description of this
theory. According to Eq.\,\eqref{ampl}, the transition amplitude
is given by 
\begin{align}
Z_{red}=\int & DA_{i}D\pi_{i}DB_{0}DB_{i}D\theta_{i}D\lambda_{a}\times\\
 & \times\sqrt{\det[M]\det[P]}\exp\left[i\int d^{4}x\left(\pi_{i}\dot{A}^{i}+\theta_{i}\dot{B}^{i}+\dot{\lambda}^{a}\Omega_{a}-{\cal V}^{(2)}\right)\right],\quad a=1,2,3\thinspace.
\end{align}
Identifying $\lambda_{1}=A_{0}$, we can write 
\begin{align}
Z_{red}= & Na^{2}\int DA_{0}DA_{i}DB_{0}DB_{i}D\pi_{i}D\theta_{i}\det\left[\left(1+a^{2}\vec{\nabla}^{2}\right)\vec{\nabla}^{2}\right]\delta\left(\left(1+a^{2}\Box\right)\vec{\nabla}\vec{A}\right)\times\nonumber \\
 & \times\delta\left(a^{2}B_{0}-\partial^{i}\theta_{i}\right)\exp\left[i\int d^{4}x\left(\pi_{i}\dot{A}^{i}+\theta_{i}\dot{B}^{i}+A_{0}\left(\partial_{i}\pi^{i}\right)-\frac{1}{a^{2}}\pi^{i}\theta_{i}-\frac{1}{2a^{4}}\theta^{i}\theta_{i}\right.\right.\nonumber \\
 & \left.\left.-\frac{1}{4}F^{ij}F_{ij}+a^{2}\partial_{i}B_{j}F^{ij}+\frac{a^{2}}{2}B_{0}B^{0}-\frac{a^{2}}{2}B_{i}B^{i}\right)\right]\thinspace,
\end{align}
Integration in $D\pi_{i}$ leads to the appearance of a delta function
$\delta\left(F^{0i}-\frac{1}{a^{2}}\theta^{i}\right)$, and further
integrations in $D\theta_{i}$ and $B_{0}$ leads to 
\begin{align}
Z_{red}= & Na^{2}\int DA_{0}DA_{i}DB_{i}\det\left[\left(1+a^{2}\vec{\nabla}^{2}\right)\vec{\nabla}^{2}\right]\delta\left((1+a^{2}\Box)\vec{\nabla}\vec{A}\right)\times\nonumber \\
 & \times\exp\left[i\int d^{4}x\left(a^{2}F_{0i}\partial^{0}{B}^{i}-\frac{1}{2}F^{0i}F_{0i}-\frac{1}{4}F^{ij}F_{ij}+\right.\right.\nonumber \\
 & \left.\left.+a^{2}\partial_{i}B_{j}F^{\ij}+\frac{a^{2}}{2}\partial_{i}F^{0i}\partial^{j}F_{0j}-\frac{a^{2}}{2}B_{i}B^{i}\right)\right].
\end{align}
Some algebraic manipulations are now in order. Up to a surface term,
we have 
\begin{equation}
a^{2}F_{0i}\partial^{0}{B}^{i}+a^{2}\partial_{i}B_{j}F^{ij}=a^{2}\partial^{\nu}F_{i\nu}B^{i}\thinspace,
\end{equation}
and completing the squares, 
\begin{equation}
a^{2}\partial^{\nu}F_{i\nu}B^{i}-\frac{a^{2}}{2}B_{i}B^{i}=-\frac{a^{2}}{2}(B^{i}+\partial_{\nu}F^{i\nu})^{2}+\frac{a^{2}}{2}\partial_{\nu}F^{i\nu}\partial^{\rho}F_{i\rho}\thinspace.
\end{equation}
Thus, by translation invariance of the functional integral, the integration
in $DB_{i}$ amounts to a $A^{\mu}$ independent Gaussian integral,
which can be incorporated in the normalization factor. As a consequence,
the transition amplitude can be cast as 
\begin{align}
Z_{red}= & N'\int DA_{0}DA_{i}\det\left[\left(1+a^{2}\vec{\nabla}^{2}\right)\vec{\nabla}^{2}\right]\delta\left((1+a^{2}\Box)\vec{\nabla}\vec{A}\right)\times\nonumber \\
 & \times\exp\left[i\int d^{4}x\thinspace\left(-\frac{1}{2}F^{0i}F_{0i}-\frac{1}{4}F^{ij}F_{ij}+\frac{a^{2}}{2}\partial_{i}F^{0i}\partial^{j}F_{0j}+\frac{a^{2}}{2}\partial_{\nu}F^{i\nu}\partial^{\rho}F_{i\rho}\right)\right]\thinspace,
\end{align}
where 
\begin{equation}
N'=Na^{2}\int DB_{i}\exp\left[-i\int d^{4}x\thinspace\frac{a^{2}}{2}(B^{i}+\partial_{\nu}F^{i\nu})^{2}\right]\thinspace.
\end{equation}
Here, we kept the seemingly dependence of $N^{\prime}$ on $A^{\mu}$
for clarity purposes. So we rewrite explicitly the following transition
amplitude in the generalized Coulomb gauge 
\begin{align}
Z_{red}=\thinspace & N'\int DA_{\mu}\det\left[\left(1+a^{2}\vec{\nabla}^{2}\right)\vec{\nabla}^{2}\right]\delta\left((1+a^{2}\Box)\vec{\nabla}\vec{A}\right)\nonumber \\
 & \times\exp\left[i\int d^{4}x\left(-\frac{1}{4}F^{\mu\nu}F_{\mu\nu}+\frac{a^{2}}{2}\partial_{\nu}F^{\mu\nu}\partial^{\rho}F_{\mu\rho}\right)\right]\nonumber \\
= & Z_{Ostro}\thinspace.
\end{align}
We therefore verity that the equivalence between the classical equations
of motion in the reduction of order and Ostrogradsky prescriptions,
seen in Eqs. (\ref{Ost1}) and (\ref{LagOstro}), hold also at the
quantum level, when we compare the transition amplitude obtained in
both prescriptions.

We end this section by making some comments to further clarify the
counting of the degrees of freedom in the Podolsky electrodynamics.
The Ostogradsky phase space has 16 variables $(\Phi^{\nu},\Gamma_{\nu},\Pi^{\mu},A_{\mu})$
and 6 constraints, so the physical phase space has 10 variables and
5 degrees of freedom \cite{Degreesfreedom}. Half of the constraints are first-class and
the other half are the gauge fixing conditions that transform the
first-class constraints into second-class constraints (for example,
imposing the generalized Coulomb gauge), such that we can determine
all the Lagrange multipliers. On other hand, in the reduced order
approach, the phase space has 16 variables $(A_{\mu},B_{\nu};\pi^{\mu},\theta^{\nu})$
and also 6 constraints, but these have different structure: two of
these constraints are second-class from the start, two are first-class
and the last two are the corresponding gauge fixing conditions. Now,
in the (FJ) methodology for the reduced order theory, we obtained
a symplectic matrix separated as Maxwell plus Proca in the first iteration
form. In doing so, the formalism already takes into account the two
second-class constraints $\Omega_{1},\Omega_{2}$ defined in Eq.\,\eqref{eq:2ndOrderConst}.
The Proca sector already presents itself as a regular sector, contributing
three degrees of freedom, while Maxwell sector is singular, and after
the introduction of two gauge fixing conditions $A_{0}=\Omega_{3}=0$
(see Eq.\,\eqref{eq:GaugeFixing}), will describe the additional
two degrees of freedom of the theory. 

\section{Final Remarks\label{sec5}}

Our main objective was to discuss the use of the (FJ) formalism for
higher derivatives theories, in particular showing how, when the order
of the equations of motion are reduced by the introduction of auxiliary
fields, the dynamics can be put in a more transparent form, with an
explicit separation of the relevant degrees of freedom.

These ideas were first presented in a toy model involving a massless
scalar field as the physical degree of freedom. We presented both
the classical and quantum basic developments of the model, both in
the Ostogradsky and the reduction of order approach, showing their
equivalence, but also pointed out that, in the latter case, one can
neatly disentangle the two degrees of freedom present in the model:
one physical massless scalar and a ghost massive one. We also briefly
discussed some recent approaches toward a consistent understanding
of these ghost fields, which present themselves as a longstanding
problem for higher derivative theories.

Afterwards, we discussed the Podolsky electrodynamics. This is a well
known higher derivative gauge theory: the (FJ) quantization procedure
have already been used for this model in the Ortogradsky formalism\,\citep{Bufalo},
while the reduced order formalism was also considered in\,\citep{Reduced}
together with the (DB) quantization procedure. We pointed out that
the combination of the reduced order with the (FJ) formalism presents
itself as a simpler way to study the constraint structure and the
quantization of this theory, since the relevant degrees of freedom
(Maxwell+Proca) are clearly separated. Our results are consistent
with the ones obtained in the other formalisms. 

It is worth noticing that Podolsky electrodynamics breaks the dual
symmetry \citep{Brandt} 
\begin{equation}
\begin{array}{l}
\vec{E}\rightarrow\vec{B}\\
\vec{B}\rightarrow-\vec{E}
\end{array}
\end{equation}
that led Dirac to consider the existence of magnetic monopoles. Hence,
a study of Podolsky equations in the vacuum may shed some light on
the question of the existence of monopoles as two Dirac strings (solenoids)
have an interaction associated with the Podolsky mass \citep{FGA2}.
Besides, the fact that the Podolsky characteristic length is associated
with the size of the electron \citep{Daniel} could lead us to explore,
by electron-positron scattering, the existence of Maxwell $\rightarrow$
Podolsky transition from the point of view of a mechanism which breaks
the dual symmetry and generate mass. These speculations derive from
our ignorance associated with the mechanisms behind the self-interaction
of the particles and their sizes and deserve rigorous scientific analysis.

Finally, we think that the natural next step of this investigation
would be the extension of this discussion for important interacting
cases (minimal coupling and sources) or the non Abelian and gravity
theories \citep{Brandt2}. These matters will be further elaborated
and requires deeper investigations.

\bigskip{}

\textbf{Acknowledgments.} This work was partially supported by Conselho
Nacional de Desenvolvimento Científico e Tecnológico (CNPq), Coordenação
de Aperfeiçoamento de Pessoal de Nível Superior (CAPES) and Fundação
de Amparo à Pesquisa do Estado de São Paulo (FAPESP), via the following
grants: FAPESP 2017/13767-9 and CNPq 304134/2017-1 (AFF), PNPD/CAPES
(AAN), CAPES PhD grant (CP).

\end{document}